\documentclass[conference]{IEEEtran}
\usepackage{cite}
\usepackage{filecontents}
\usepackage{stfloats}
\usepackage{color}

\newcommand{\tunderbrace}[1]{\underbrace{\scriptstyle#1}}

\definecolor{ao}{rgb}{0.0, 0.5, 0.0}
\definecolor{copper}{rgb}{0.72, 0.45, 0.2}
\ifCLASSINFOpdf
  \usepackage[pdftex]{graphicx}
  \graphicspath{{images/}}
  \DeclareGraphicsExtensions{.pdf,.jpeg,.png}
\else
  \usepackage[dvips]{graphicx}
  \graphicspath{{images/}}
  \DeclareGraphicsExtensions{.eps}
\fi

\usepackage[cmex10]{amsmath}

\usepackage{array}
\usepackage{mdwmath}
\usepackage{mdwtab}

\usepackage[tight,footnotesize]{subfigure}
\usepackage{url}
\newtheorem{theorem}{Theorem}
\newtheorem{corollary}{Corollary}
\newtheorem{Lemma}{Lemma}

\newcommand{\goodchi}{\protect\raisebox{2pt}{$\chi$}}
\usepackage{amssymb}
\usepackage{bm}
\usepackage{mathrsfs}
\newtheorem{remark}{Remark}
\newtheorem{definition}{Definition}

\newtheorem{conjecture}{Conjecture}
\newtheorem{proposition}{Proposition}
\begin{document}
\title{Generalized Interlinked Cycle Cover\\ for Index Coding}

\author{\IEEEauthorblockN{Chandra Thapa, Lawrence Ong, and Sarah J. Johnson}
\IEEEauthorblockA{School of Electrical Engineering and Computer Science, The University of Newcastle, Newcastle, Australia\\
Email: chandra.thapa@uon.edu.au, lawrence.ong@cantab.net, sarah.johnson@newcastle.edu.au} %
}
\maketitle
\begin{abstract}
A source coding problem over a noiseless broadcast channel where the source is preinformed about the contents of the cache of all receivers, is an index coding problem. Furthermore, if each message is requested by one receiver, then we call this an index coding problem with a unicast message setting. This problem can be represented by a directed graph. 
In this paper, we first define a structure (we call generalized interlinked cycle ($ \mathsf{GIC} $)) in directed graphs. A $ \mathsf{GIC} $ consists of cycles which are interlinked in some manner (i.e., not disjoint), and it turns out that the $ \mathsf{GIC} $ is a generalization of cliques and cycles. We then propose a simple scalar linear encoding scheme with linear time encoding complexity. This scheme exploits $ \mathsf{GICs} $ in the digraph. We prove that our scheme is optimal for a class of digraphs with message packets of any length. Moreover, we show that our scheme can outperform existing techniques, e.g., partial clique cover, local chromatic number, composite-coding, and interlinked cycle cover.  
\end{abstract}
\begin{IEEEkeywords}
Index coding problem, unicast, optimal broadcast rate, linear codes, interlinked cycles.
\end{IEEEkeywords}
%
\IEEEpeerreviewmaketitle
\section{Introduction}

We consider a transmitter broadcasting message through a noiseless broadcast channel to multiple receivers, each knowing some message packets a priori, which is known as \emph{side information}. The \emph{side information} of receivers can be utilized in order to reduce the number of coded symbols required to be broadcast by the transmitter to all receivers. This is known as the index coding problem, and was introduced by Birk and Kol in 1998 \cite{ISCOD}. To date, the index coding problem is an open problem. The main aim of the index coding problem is to find an index code that has the minimum number of coded symbols. 

When each message is requested by one receiver (i.e., unicast), the index coding problems can be modelled by digraphs (i.e., directed graphs). This paper considers the unicast message setting. 

Linear index codes (scalar and vector linear) \cite{ISCOD,maisbound,linearcodeoptimal,chaudhary,eden} have simpler encoding and decoding process than non-linear index codes. In the literature, optimal scalar linear index codes can be characterized by a graph function called the \emph{minrank} function \cite{maisbound}. However, finding \emph{minrank} for a general digraph is NP-hard \cite{NPhard1}, and does not provide much intuition on the interaction between the side information configuration and the index codes. Thus in this paper, we use the graph-theoretic approach to exploit specific graph structures. 

There are various graph-theoretic approaches such as clique cover \cite{ISCOD}, partial clique cover \cite{ISCOD}, cycle cover \cite{maisbound, chaudhary,neely}, and graph coloring (including fractional) \cite{ISCOD, linearprogramming, localgarphcoloring}, which exploit the graph structure during encoding of the messages to \emph{save} transmissions (i.e., compared to sending uncoded message packets). In our earlier work \cite{ourpaper}, we presented a new coding scheme exploiting interlinked cycles in digraphs. This new scheme, called interlinked cycle cover ($ \mathsf{ICC} $), generalized the clique cover and the cycle cover schemes. We proved that for a class of digraphs, called $ \mathsf{ICC} $ digraphs, the $ \mathsf{ICC} $ scheme is optimal. Furthermore, we showed that for some examples, it can outperform some existing schemes. In an interlinked cycle structure with $ N $ number of vertices, there exists a set of $ K $ vertices, where each vertex has a directed path to every other vertex of the set. Each of these $ K $ vertices must have an out-degree equal to $ K-1 $, all remaining vertices (i.e., $ N-K $ vertices) must have an out-degree equal to one, and both can have an in-degree greater than or equal to one. These conditions on out-degree of vertices make the definition of interlinked cycles rather limiting, and it restricts the size of the class of $\mathsf{ICC} $ digraphs (for which the $\mathsf{ICC}$ scheme is optimal). Moreover, we were unable to show that the $ \mathsf{ICC} $ scheme can outperform the composite-coding scheme (based on random coding approach which requires infinitely long message packets) proposed by Arbabjolfai et al. \cite{composite}.   

\subsection{Our Contributions}
We first redefine (and extend) the previous definition of the interlinked cycle structure, so that both the in-degree and the out-degree of any vertex in it are allowed to be greater than or equal to one. The resultant interlinked cycle structure is called a generalized interlinked cycle ($ \mathsf{GIC} $). We then propose a simple encoding scheme based on the $ \mathsf{GIC} $, called the generalized interlinked cycle cover ($ \mathsf{GICC} $). The $ \mathsf{GICC} $ scheme generalizes the $ \mathsf{ICC} $ scheme, the clique cover scheme, and the cycle cover scheme. Furthermore, we characterize a class of digraphs where the $ \mathsf{GICC} $ is optimal (over all codes, including non-linear index codes), and show that the $ \mathsf{GICC} $ scheme can outperform existing techniques (including partial clique cover, local chromatic number, composite-coding, and interlinked cycle cover).   
\section{Definitions}
Consider a transmitter that wants to transmit $ N $ message packets $ X = \{x_1,x_2,\dotsc,x_N\} $ to $ N $ receivers $ \{1,2,\dotsc,N\} $ in a unicast message setting such that each receiver $ i $ is requesting a message packet $ x_i $. Moreover, each receiver $ i $ has side information $ S_i\subseteq X\setminus \{x_i\} $. This problem can be described by a digraph $D =(V(D),A(D))$, where $ V(D)=\{1,2,\dotsc,N\} $ is a set of vertices representing the $ N$ receivers. An arc $ (i\rightarrow j) \in A (D) $ exists from vertex $ i $ to vertex $ j$ if and only if receiver $ i $ has packet $ x_j $ (requested by receiver $ j $) as its side information. The side information of vertex $ i $ is $ S_i \triangleq \{x_j: j\in N_D^+(i)\} $, where $ N_D^+ (i) $ is the out-neighborhood of $ i $ in $ D $. For simplicity, we use the term ``messages" to refer to message packets in the remainder of this paper. 
\begin{definition} [Valid index code]
Suppose $ x_i \in \{0,1\}^t $ for all $ i $, for some integer $ t\geq 1 $, i.e., each message consists of $t$ bits. 
Given an index coding problem described by $D$, a valid index code ($ \mathscr{F} $,$ \{\mathscr{G}_i\} $) is defined as follows:
\begin{enumerate}
\item An encoding function for the source, $ \mathscr{F}:\{0,1\}^{Nt} \rightarrow \{0,1\}^p $, which maps $X$ to a $ p $-bit index for some integer $ p $. 
\item A decoding function $ \mathscr{G}_i $ for every receiver $ v_i $, $ \mathscr{G}_i: \{0,1\}^p \times \{0,1\}^{|S_i|t}\rightarrow \{0,1\}^t $, that maps the received index $ \mathscr{F}(X) $ and its side information $ S_i $ to the requested  message $ x_i $.   
\end{enumerate}  

 The broadcast rate of the ($ \mathscr{F},\{\mathscr{G}_i\} $) index code is the number of transmitted bits per received message bits at every user, or equivalently the number of coded packets (of $t$ bits), and this is denoted by $ \ell(D) \triangleq \frac{p}{t}$. Thus the optimal broadcast rate for a given index coding problem $D$ with $ t $-bit message is $ \beta_t(D)\triangleq \underset{\mathscr{F}}{\mathrm{min}}\ \ell(D)$, and the optimal broadcast rate over all $ t $ is defined as $ \beta(D)\triangleq \underset{t}{\mathrm{inf}}\ \beta_t(D)$.
 \end{definition}
\begin{definition} [Path and cycle]
A \emph{path} contains a sequence of unique (except possibly the first and last) vertices, say $1,2,\dotsc,M $, and an arc $( i \rightarrow i+1 )$ for each consecutive pair of vertices $ (i, i+1) $ for all $ i\in \{1,\dotsc,M-1\} $. We represent a path in a digraph $ D $ as $ P_{1\rightarrow M}(D)=\langle 1,\dotsc,M \rangle $. Here, $1$ is 
the \emph{first vertex}, $ M $ is the \emph{last vertex}, and all remaining vertices ($ 2,3,\dotsc,M-1 $) are \emph{internal vertices} of the path. A path with the same first and last vertex is a \emph{cycle}. 
\end{definition}  

	
	\section{The $\mathsf{GIC}$ structure and Code construction}
	\subsection{Description of a $ K $-$\mathsf{GIC}$ structure}
	\begin{figure}[t] 
		\centering
			\subfigure []{ %
		        \includegraphics[height=4.6cm,keepaspectratio]{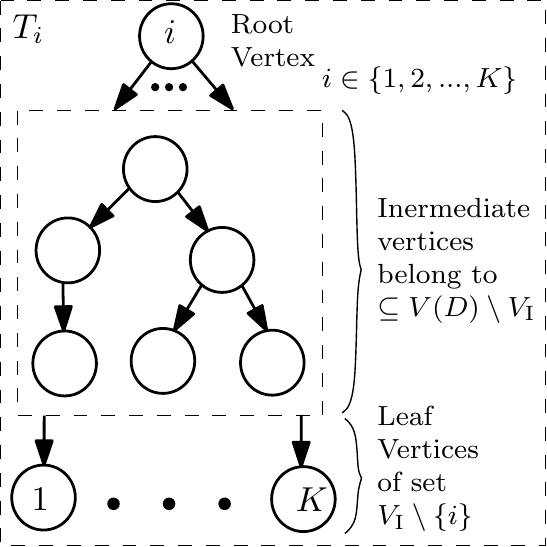}
		        \label{fig1}}
		        \hskip2pt	
		     \subfigure []{
		        \centering
		        \includegraphics[height=4.8cm,keepaspectratio]{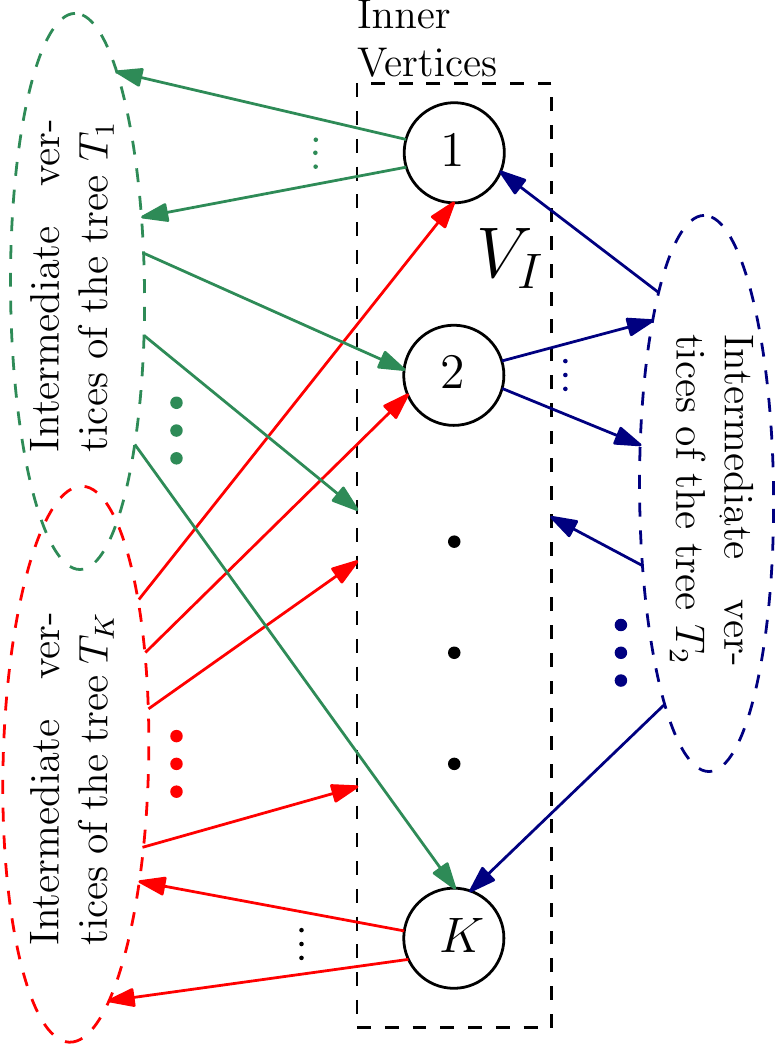}
		        \label{fig2}
			}
		\caption{(a) Outline of a tree $ T_i$, which has the root vertex $ i \in  V_{\mathrm{I}}$, all vertices in $ V_{\mathrm{I}} \setminus \{i\} $ as the leaf vertices, and some or all vertices in $ V(D)\setminus V_{\mathrm{I}} $ (i.e., non-inner vertex set) as intermediate vertices (between the root and the leaves), and b) outline of the structure of a sub-digraph $ D_{\mathrm{s}} $, where different trees $ T_i $ (shown in different colors) can share non-inner vertices.}  
		\label{fig}
	
	\end{figure}
	
	Consider a graph structure with $ N $ vertices having the following properties:
	\begin{enumerate}
	\item A set of $ K $ vertices, denoted by $ V_{\mathrm{I}} $, such that for any ordered pair of its vertices ($ i,j $), $ i\neq j $, there is a path from $ i $ to $ j $ which does not include any other vertex of $ V_{\mathrm{I}} $. We call $ V_{\mathrm{I}} $ the \emph{inner vertex set}, and without loss of generality, we represent it as $ V_{\mathrm{I}}=\{1,2,\dotsc,K\} $. The vertices of $ V_{\mathrm{I}}$ are refereed to as \emph{inner vertices}. 
	\item Due to the existence of paths in between any $ (i,j) \in V_{\mathrm{I}} $, for each vertex $ i$, we can always find a \emph{directed rooted tree} in $ D $, denoted by $ T_i $ where vertex $ i $ is the root vertex, and all other vertices $V_{\mathrm{I}} \setminus \{i\} $ are the leaves (see Fig.\ \ref{fig1}). The trees may be non-unique.
	\end{enumerate}
	Denote the union of all selected $ K $ trees as  $ D_{\mathrm{s}}\triangleq  \bigcup_{\forall i\in {V_{\mathrm{I}}}} T_i $ (see Fig. \ref{fig2}). If $D_{\mathrm{s}} $ satisfies two conditions (to be defined shortly), we call it a $ K $-$ \mathsf{GIC} $ structure (denoted as a $ K $-$ \mathsf{GIC} $ sub-digraph: $ D_K=(V(D_K),A(D_K)) $, where  $ |V(D_K)|=N $ and $ V_{\mathrm{I}} = \{1,2,\dotsc,K\} $). Now we define a type of cycle and a type of path.
	\begin{definition} [I-cycle]
	A cycle that includes only one inner vertex $ i\in V_{\mathrm{I}} $ is an \emph{I-cycle}. 
	\end{definition}
	\begin{definition} [P-path]
	A path in which only the first and the last vertices are from $ V_{\mathrm{I}} $, and they are distinct, is a \emph{P-path}. 
	\end{definition}
	
	The conditions for a $ D_{\mathsf{s}} $ to be qualified as a $\mathsf{GIC}$ are as follows:
	\begin{enumerate}
		\item \emph{Condition 1:} There is no \emph{I-cycle}.
		\item \emph{Condition 2:} For all ordered pairs of inner vertices ($ i,j $), $ i\neq j $, there is only one \emph{P-path} from $ i $ to $ j $. 
	\end{enumerate}
	These two conditions are necessary for our code construction described in the following section. 
	\vskip-5pt
	\subsection{Code construction for a $ K $-$ \mathsf{GIC} $ structure} \label{secb}
	\vskip-2pt
	We propose the following coded symbols (which form a scalar linear index code) for a $ K $-$ \mathsf{GIC} $ sub-digraph $ D_K $:
	\begin{enumerate}
	\item A coded symbol obtained by the bitwise XOR (denoted by $ \oplus $) of messages (each of $ t $-bits) requested by all vertices of the inner vertex set $ V_{\mathrm{I}} $, i.e.,
	\begin{equation}\label{innervertexcodesymbol}
	w_{\mathrm{I}} \triangleq \bigoplus\limits_{i=1}^K x_i.
	\end{equation}
	\item For each non-inner vertex $ j \in \{K+1,K+2,\dotsc,N\}$, a coded symbol obtained by the bitwise XOR of the message requested by $ j $ with the messages requested by its out-neighborhood vertices, i.e.,
	\begin{equation}\label{outervertexcodesymbol}
	w_j\triangleq x_j\oplus \bigoplus\limits_{q\ \in\ N^+_{D_K}(j)} x_q.
	\end{equation}
	\end{enumerate} 
	Denote this index code constructed for the $ K $-$ \mathsf{GIC} $ sub-digraph by $ W\triangleq~ \{w_{\mathrm{I}},w_j:K+1\leq j\leq N\} $. The calculation of the total number of coded symbols, each of $ t$-bits, in $ W $ is straightforward,
	\begin{equation} \label{codelength}
	\ell (D_K)=N-K+1.
	\end{equation}
	\begin{remark}
		Given a $ K $-$ \mathsf{GIC} $ sub-digraph $ D_K $, encoding $ W $ requires at most $ t \times \left(  (K-1)  +\sum\limits_{i\in V(D_K)\setminus V_{\mathrm{I}}}^{} |N^+_{D_K}(i)| \right) $ bit-wise XOR operations.
	\end{remark}
	
	Now we show that all $ N $ vertices in $ D_K $ can decode their respective requested messages from $ W $.
	
	From \eqref{outervertexcodesymbol}, all $j \in  \{K+1,K+2,\dotsc,N\} $ which are non-inner vertices, can decode their requested messages. This is because the coded symbol $ w_j $ is the bitwise XOR of the messages requested by $ j $ and its all out-neighborhood vertices, and any $ j $ knows messages requested by all of its out-neighborhood vertices as side information. 
		
	For an inner vertex $ i $, rather than analyzing the sub-digraph $ D_K $, we will analyze its tree $ T_i$, and show that it can decode its message from the relevant symbols in $ W $. We are able to consider only the tree $ T_i $ due to the following proposition.
	\begin{proposition} \label{lemma3}
		If a vertex $ v \in V(T_i) $ such that $ v \notin V_{\mathrm{I}}$, then the out-neighborhood is the same in the tree $ T_i $ and in $ D_K $, i.e., $N^+_{T_i}(v)=N^+_{D_K}(v) $. 	
	\end{proposition}
	\begin{IEEEproof}
		Refer to Appendix \ref{append1}.
	\end{IEEEproof}	 
	
	Now let us take any tree $ T_i $. Assume that it has a height $ H $ where $ 1\leq H\leq (N-K+1) $. The vertices in $ T_i $ are at various depths, i.e., $ \{0,1,2,\dotsc, H\} $ from the root vertex $ i $. The root vertex $ i $ has depth zero, and any vertex at depth equal to the height of the tree is a leaf vertex.  
	
	First of all, in the tree $ T_i $, we compute the bitwise XOR among coded symbols of all non-leaf vertices at depth greater than zero, i.e., $ Z_i \triangleq \bigoplus_{j\in V(T_i)\setminus V_{\mathrm{I}}} w_j $. However, in the tree $ T_i $, the message requested by a non-leaf vertex, say $p$, at a depth greater than one, appears exactly twice in $ \{w_j:j\in V(T_i)\setminus V_{\mathrm{I}} \} $; 
	\begin{enumerate}
		\item [i)] once in $ w_k $, where $ k $ is parent of $ p $ in tree $ T_i $, and
		\item [ii)] once in $ w_p $.
	\end{enumerate}
	Thus they cancel out each other while computing $ Z_i $ in the tree $ T_i $ (see \eqref{eqB} in Appendix \ref{append2}). Hence, in the tree $ T_i $, the resultant expression which is bitwise XOR of 
	\begin{enumerate}
			\item [i)] messages requested by all non-leaf vertices at depth one, and
			\item [ii)] messages requested by all leaf vertices at depth greater than one, 
	\end{enumerate} 	
	is obtained (see \eqref{eqD} in Appendix \ref{append2}). Secondly, in the tree $ T_i $, we compute $ w_{\mathrm{I}} \oplus Z_i$ (see \eqref{xvt2} in Appendix \ref{append2}) which yields the bitwise XOR of 
	\begin{enumerate}
			\item [i)] messages requested by all non-leaf vertices at depth one which are in out-neighborhood of $ i $,
			\item [ii)] messages requested by all leaf vertices at depth one which are also in out-neighborhood of $ i $, and
			\item [iii)] message requested by $ i $, i.e., $ x_i $. 
	\end{enumerate} 	
	This is because the messages requested by each leaf vertex at depth greater than one in the tree $ T_i $ is present in both the resultant terms of $ Z_i $ and in $ w_{\mathrm{I}}  $, thereby they cancel out itself in $ w_{\mathrm{I}} \oplus Z_i$. Hence, $ w_{\mathrm{I}} \oplus Z_i$ yields the bitwise XOR of $ x_i $ and $\{x_j:j \in N^+_{D_K}(i)\} $. As $ i $ knows all $\{x_j:j \in N^+_{D_K}(i)\} $ as side-information, any inner vertex $ i $ can decode its required message from $ w_{\mathrm{I}} \oplus Z_i$.     
		

	\section{Results}

    \begin{definition} [Generalized interlinked cycle cover $(\mathsf{GICC})$ scheme]
	For any digraph, the $ \mathsf{GICC} $ scheme finds a set of disjoint $ \mathsf{GIC} $ sub-digraphs. It then (a) codes each of these $ \mathsf{GIC} $ sub-digraphs using the code construction described in Section \ref{secb}, and (b) sends uncoded messages requested by all remaining vertices (i.e., vertices which are not in any of these disjoint $ \mathsf{GIC} $ sub-digraphs).  
	\end{definition}

        Now we present a main result of this paper. It is best expressed in terms of \textit{savings} defined as follows:
	
    \begin{definition} [Savings]
	The number of packets saved (i.e., $ N-\ell(D)$), by transmitting coded symbols (coded packets) rather than transmitting uncoded message packets, is called \emph{savings}. 
	\end{definition}

	\begin{theorem} \label{theorem1}
	For any digraph D, a valid index code of length $ \ell_{\mathsf{GICC}}(D)=N-\sum\limits_{i=1}^{\psi} (K_i-1)$ can be achieved by using the $ \mathsf{GICC} $ scheme, where ($ K_i-1 $) is the saving in each disjoint $ K_i $-$ \mathsf{GIC} $ sub-digraph, and $ \psi $ is the number of disjoint $ \mathsf{GIC} $ sub-digraphs in $ D $.
	\end{theorem}
\begin{IEEEproof}
	Consider a $ K $-$ \mathsf{GIC} $ sub-digraph $ D_K $. It follows from \eqref{codelength} that the total number of savings achieved by the $ \mathsf{GICC} $ scheme is 
	\begin{equation} \label{eq}
	N-\ell_{\mathsf{GICC}}(D_K)=N-(N-K+1)=K-1.  
	\end{equation}
	
	For any digraph $ D $ containing $ \psi $ disjoint $ \mathsf{GIC} $ sub-digraphs, a saving of $ K_i-1 $ is obtained in each  $ D_{K_i}$ (from \eqref{eq}), where $ i\in \{1,\dotsc,\psi\} $. Now the total saving is the summation of savings in all disjoint $ \mathsf{GIC} $ sub-digraphs, i.e., $ \sum_{i=1}^{\psi}(K_i-1) $. Hence, $ \ell_{\mathsf{GICC}}(D)=N-\sum_{i=1}^{\psi} (K_i-1)$.   
\end{IEEEproof}
\begin{remark}
	The  $ \mathsf{GIC} $ sub-digraphs found by the  $ \mathsf{GICC} $ scheme are not unique. So, finding the best $ \ell_{\mathsf{GICC}}(D) $ involves optimizing over all choices of disjoint $ \mathsf{GIC} $ sub-digraphs in $ D $, and this requires high time complexity. We will leave the design of algorithms or approximations as our future work.   
\end{remark}

	\subsection{$ \mathsf{GIC}$ sub-digraphs include $ \mathsf{ICC}$ sub-digraphs}
	\begin{theorem}
	$ \mathsf{GIC} $ sub-digraphs include $ \mathsf{ICC} $ sub-digraphs as a special case. 
	\end{theorem}
	\begin{IEEEproof}
	An $ \mathsf{ICC} $ sub-digraph $ D_{\mathsf{ICC}} $, is defined as follows \cite{ourpaper}:
	\begin{enumerate}
	\item \label{item:path-i} It has $ k $ disjoint paths, $ P_i \triangleq\langle v_1^i,v_2^i,\dotsc,v_{n_i}^i \rangle$, for each  $ i \in \{1,2,\dotsc,k\} $, where each $ P_i $ has $ n_i \geq 1 $ number of vertices.
    \item  \label{item:path} For any distinct $i,j \in \{1,2,\dotsc,k\}$, there is a path from $v_{n_i}^i \in V(P_i)$ to some $v^j \in V(P_j) $, denoted as $ \langle v_{n_i}^i,v_1^{ij},\dotsc,v_{n_{ij}}^{ij},v^j \rangle $. Denote the sub-path  $ P_{i,j} \triangleq\langle v_1^{ij},v_2^{ij},\dotsc,v_{n_{ij}}^{ij} \rangle$, where each $ P_{i,j} $ has $ n_{ij} \geq 0 $ number of vertices.
	\item The set of vertices in all $ P_i $ and in all $ P_{i,j} $ are mutually disjoint.
	\item \label{item:in} Each first vertex $ v_1^j $ in $ P_j $ has at least one in-degree.  
 	\end{enumerate}  

    Select $ V_\mathrm{I}=\{v_{n_1}^1,v_{n_2}^2,\dotsc,v_{n_k}^k\} $. Point~\ref{item:path} guarantees that, for any ordered pair ($ i,j $), $ i\neq j $ and $ i,j \in \{1,2,\dotsc, k\} $, there exists a path from $ v_{n_i}^i $ to $ v_{n_j}^j $. Now we will show that the $ \mathsf{ICC} $ sub-digraph with the chosen $ V_\mathrm{I} $, satisfies the two conditions to be a $ \mathsf{GIC}$ sub-digraph $ D_K $.

	By construction of an $ \mathsf{ICC} $ digraph, we have the following:
	For any $ v_{n_i}^i \in V_\mathrm{I} $, all paths from $ v_{n_i}^i $ must go through some $ v^j \in V(P_j) $ for some $ j $, and then $ v_{n_j}^j $ before returning to $ v_{n_i}^i $. Therefore, there is no $ I $-cycle. 

    Note that each vertex in $V(D_\mathsf{ICC}) \setminus V_\mathrm{I}$ has out-degree one. Thus it follows from points~\ref{item:path-i} and~\ref{item:path} that there is only one $P$-path between any  $i,j \in V_\mathrm{I}$.
	    
	Since there is only one $P$-path from $ i $ to $ j $ for any $i, j \in V_\mathrm{I}$, every $ P_{i,j} $, which is part of the $ P $-path from $ i $ to $ j $, must be a sub-digraph of the tree $ T_i $. For every $ j \in V_\mathrm{I} $, point 4 dictates that the entire $ P_j $ must be the part of the $ P$-path from some $ i $ to $ j $. Hence, $ P_j $ is a sub-digraph of the tree $ T_i $. We have included all vertices and arcs in the construction of the trees in $D_K$. Consequently, $ D_\mathsf{ICC}=D_K$.
	\end{IEEEproof}	
	\subsection{The $ \mathsf{GICC}$ scheme includes the $ \mathsf{ICC}$ scheme as a special case}
	\begin{theorem}
	The $ \mathsf{GICC}$ scheme includes the $ \mathsf{ICC}$ scheme.
	\end{theorem}
	\begin{IEEEproof}
	In an $ \mathsf{ICC} $ sub-digraph $ D_{\mathsf{ICC}} $ \cite{ourpaper}, select $ V_\mathrm{I}=\{v_{n_1}^1,v_{n_2}^2,\dotsc,v_{n_k}^k\} $. The coded symbols $ w'=\bigoplus_{i=1}^{k} x_{n_i}^i$ of the $ \mathsf{ICC} $ is the same as the coded symbol $w_{\mathrm{I}}$ of the $ \mathsf{GICC} $. Now for the remaining vertices (of set $V(D_{\mathrm{ICC}})\setminus V_{\mathrm{I}}$), the coded symbols of the $ \mathsf{ICC} $ scheme are simply the bitwise XOR of the messages requested by each vertex $ j $ and its out-neighborhood vertices, which is same as $w_{j}$, $ j\in V(D_K)\setminus V_{\mathrm{I}}$ of the $ \mathsf{GICC} $ scheme.
	\end{IEEEproof}
	
	\subsection{The $ \mathsf{GICC}$ scheme includes the cycle cover and clique cover schemes as special cases}

	\begin{corollary}
	The $ \mathsf{GICC} $ scheme includes the cycle cover and the clique cover schemes as special cases.
	\end{corollary}
	\begin{IEEEproof}
	The $ \mathsf{ICC} $ scheme includes the cycle cover scheme and the clique cover scheme as its special cases \cite{ourpaper}. From Theorem 2, the $ \mathsf{GICC} $ scheme includes $ \mathsf{ICC} $ scheme as its special case.   
	\end{IEEEproof}
	\subsection{The $ \mathsf{GICC} $ scheme is optimal for a class of digraphs} \label{sec:}
	We first prove a lemma that will help to prove the optimality of the $ \mathsf{GICC} $ scheme.
	
		\begin{Lemma} \label{lemma6}
			In a $ \mathsf{GIC} $ sub-digraph, any cycle must contain either (i) no inner vertex, or (ii) at least two inner vertices. 
		\end{Lemma}
		\begin{IEEEproof}
			It follows directly from the property of a $ \mathsf{GIC} $ sub-digraph that a cycle cannot be formed by including only one inner vertex because this type of cycle is an $ I $-cycle. 
		\end{IEEEproof}
		\begin{definition} [Maximum acyclic induced subgraph (\emph{MAIS})]
			For a digraph $ D $, an induced acyclic sub-digraph with the largest number of vertices is called the MAIS. We denote the order of an MAIS by $\mathsf{MAIS}(D) $.
		\end{definition}
		
		It has been shown~\cite{maisbound} that for any $D$ and $t$,
		\begin{equation}
		\mathsf{MAIS}(D) \leq \beta(D) \leq \beta_t(D) \leq \ell(D). \label{eq:lower-bound}
		\end{equation}

		\begin{theorem} \label{th3}
			For a class of digraphs, in which each digraph is a $K$-$\mathsf{GIC}$ sub-digraph $ D_K $ with 
			\begin{itemize}
				\item (Case 1) no cycle among the non-inner vertices, or
				\item (Case 2) $ M\geq 1 $ disjoint cycles among non-inner vertices, and we can group the inner vertex set $ V_{\mathrm{I}} $ in to $ M+1 $ sub-sets such that each of them forms a disjoint $ \mathsf{GIC} $ sub-digraph of case 1, and such $ \mathsf{GIC} $ sub-digraphs are also disjoint from the $ M $ cycles among non-inner vertices,   
			\end{itemize}
			the scalar linear index code given by the $ \mathsf{GICC} $ scheme is optimal for messages of any $ t\geq1 $ bits, i.e., $ \ell_{\mathsf{GICC}}(D_K)=\beta(D_K) = \beta_t(D_K)$.  
		\end{theorem}
		\begin{IEEEproof}
			We will show that the MAIS lower bound \eqref{eq:lower-bound} is tight for all $t$. We denote the digraph by $D_K$, and consider that it has $ N $ number of vertices. For $K=1$, the digraph contains only one vertex, and $\mathsf{MAIS}(D_1) = 1$. For $K \geq 2$, we have the following:
			
			(Case 1) From Lemma \ref{lemma6}, any cycle must include at least two inner vertices, or no inner vertex, thus if we remove $ K-1 $ inner vertices, then the digraph $ D_K $ becomes acyclic. Thus
			\begin{equation} \label{upperboundmais1}
			\mathsf{MAIS}(D_K) \geq N-K+1.
			\end{equation}
			From Theorem \ref{theorem1}, we get 
			\begin{equation} \label{lowerboundmais1}
			\ell_{\mathsf{GICC}}(D_K)=N-K+1.
			\end{equation}
			It follows from \eqref{eq:lower-bound}, \eqref{upperboundmais1}, and \eqref{lowerboundmais1} that $ \mathsf{MAIS}(D_K)=N-~K+~1=\ell_{\mathsf{GICC}}(D_K)$. Thus $\ell_{\mathsf{GICC}}(D_K) =\beta(D_K) = \beta_t(D_K) = N-K+1$.
			
			(Case 2) A $ D_K$ can be viewed in two ways. The first way is considering the whole $ D_K $ as a $ K $-$ \mathsf{GIC} $ digraph. The second way is considering induced sub-digraphs of $ D_K$ which consist of; (a) $ M $ disjoint cycles together consisting of a total of $ N_A $ ($0\leq N_A < N-K $) non-inner vertices (if $ N_A=0$ or $ 1 $, then $ M=0$, which is case 1), (b) $ M+1 $ disjoint $ \mathsf{GIC} $ sub-digraphs each with $ N_i $ number of vertices and $ K_i$ number of inner vertices in such a way that $\sum_{i=1}^{M+1} K_i=K$, we consider that each $ \mathsf{GIC} $ sub-digraph is also disjoint from all $ M $ cycles among non-inner vertices, and (c) total remaining of $ N_B =N-N_A-\sum_{i=1}^{M+1} N_i $ non-inner vertices (which are not included in $ M $ cycles, or the $ M+1 $ $ \mathsf{GIC} $ sub-digraphs).
			Now we will show that both ways of looking at $ D_K $ are equivalent in the sense of the index code length generated from our proposed scheme, and both equal to $ \mathsf{MAIS}(D_K)$. We prefer the second way of viewing $ D_K $ for our proof since it is easier to find the MAIS lower bound.   
			
			For the partitioned $ D_K $ (looking at in the second way), the total number of coded symbols will be the summation of the length of the coded symbols for (i) each of the $ M $ disjoint cycles (each cycle has saving equal to one), (ii) each of the $ M+1 $ disjoint $ \mathsf{GIC} $ sub-digraphs (each of $ \mathsf{GIC} $ sub-digraphs has savings equal to $ K-1 $), and (iii) $ N_B $ uncoded symbols for the remaining non-inner vertices, i.e.,  
			\begin{align} \label{EQL2}
			\ell_{\mathsf{GICC}}^{'} (D_K) &= (N_A-M)+ \sum\limits_{i=1}^{M+1}(N_i-K_i+1)+N_B\nonumber  \\
			&= N- K+ 1. 
			\end{align}    
			From \eqref{lowerboundmais1} and \eqref{EQL2}, $ \ell_{\mathsf{GICC}}(D_K)=\ell_{\mathsf{GICC}}^{'} (D_K) $, thus from both perspectives the code length is the same. 
			
			Now for $ D_K $ (looking at in our second way), if we remove one vertex from each of the $ M $ cycles among non-inner vertices ($ M $ removal in total), and remove $ K_i-1 $ vertices from each of the $ M+1 $ $ \mathsf{GIC} $ sub-digraphs ($ \sum_{i=1}^{M+1}(K_i-1)=K-M-1$), i.e., total removal of $ K-1 $, then the digraph becomes acyclic. Thus
			\begin{equation} \label{upperboundmais2}
			\mathsf{MAIS}(D_K) \geq (N-K+1).
			\end{equation}
			It follows from \eqref{eq:lower-bound}, \eqref{lowerboundmais1}, and \eqref{upperboundmais2} that $ \mathsf{MAIS}(D_K)=N-~K+~1=\ell_{\mathsf{GICC}}(D_K)$. Thus $\ell_{\mathsf{GICC}}(D_K) =\beta(D_K) = \beta_t(D_K) = N-K+1$.
		\end{IEEEproof}
		
		\begin{conjecture}
			For any $ K $-$ \mathsf{GIC} $ digraph $ D_K $, and message of any $ t\geq 1 $ bits, the scalar index codes given by $ \mathsf{GICC} $ scheme is optimal, i.e., $ \ell_{\mathsf{GICC}}(D_K)=\beta(D_K) = \beta_t(D_K)$.  
		\end{conjecture}
		
		If $ \mathsf{MAIS}(D_K) < \ell_{\mathsf{GICC}}(D_K) $, we conjecture that we can always find disjoint $ \mathsf{GIC} $ sub-digraphs such that the summation of all number of coded symbols required for each  $ \mathsf{GIC} $ sub-digraph is equal to $ \mathsf{MAIS}(D_K)$.  

\section{Comparison with Existing Techniques}

\subsection{The $\mathsf{GICC} $ scheme can outperform existing techniques including composite-coding and interlinked cycle cover}

The $\mathsf{GIC} $ digraph shown in Fig. \ref{4a} has $ K=4 $ and is denoted by $ D_{4}$. The coded symbols from the $\mathsf{GICC} $ scheme have length $ \ell_{\mathsf{GICC}}(D_{4})=3$, and they are $\{1\oplus 2\oplus 3\oplus 4 ,\ 5 \oplus 2 \oplus 3,\ 6\oplus 3 \oplus 4\}$. The upper bounds for the digraph $ D_{4} $ by existing techniques are (a) composite-coding \cite{composite}, $ \ell_{\mathsf{CC}}(D_{4})=3.5 $; (b) local chromatic number \cite{localgarphcoloring}, $ \ell_{\mathsf{LC}}(D_{4})=4$; (c) fractional partial clique cover \cite{localtimesharing}, $\ell_{\mathsf{FPCC}} (D_{4})=4 $; (d) interlinked cycle cover \cite{ourpaper}, $ \ell_{\mathsf{ICC}} (D_4)=4 $; (e) clique cover \cite{ISCOD}, $ \ell_{\mathsf{CL}} (D_4)=5$; and (f) cycle cover \cite{maisbound, chaudhary,neely}, $ \ell_{\mathsf{CY}} (D_4)=4$. All of these upper bounds are strictly greater than $ \ell_{\mathsf{GICC}}(D_{4})$.

\vskip-5pt
\subsection{Performance improvement of the $ \mathsf{GIC} $ scheme over existing schemes}

We outline a class of $ \mathsf{GIC} $ digraphs with $ K $ number of inner vertices, and $ 2(K-1) $ number of non-inner vertices, so that $ N=3K-2 $. Without loss of generality, say $ V_{\mathrm{I}}=\{1,2,\dotsc,K\} $, and the non-inner vertex set is $\{K+1,\dotsc,N\}  $. Now for $ i\in \{2,\dotsc,K-1\} $, every vertex $ i $ knows messages requested by vertices $ K+i $ and $ 3K-i $. Vertex $K+i$ knows all messages requested by vertices $\{i+1,\dotsc,K\}$, and vertex $ 3K-i $ knows all messages requested by vertices $ \{1,\dotsc,i-1\} $. The first inner vertex (i.e., vertex 1) knows the message requested by vertex $ K+1 $, and vertex $ K+1 $ knows all messages requested by vertices $ V_{\mathrm{I}}\setminus \{1\} $. Similarly, the last inner vertex (i.e., vertex $ K $) knows the message requested by vertex $ 2K$, and vertex $ 2K $ knows all messages requested by vertices $ V_{\mathrm{I}}\setminus \{K\} $. 

\begin{figure}[t] 
	\centering
	\subfigure []{ %
		\includegraphics[height=3.2cm,keepaspectratio]{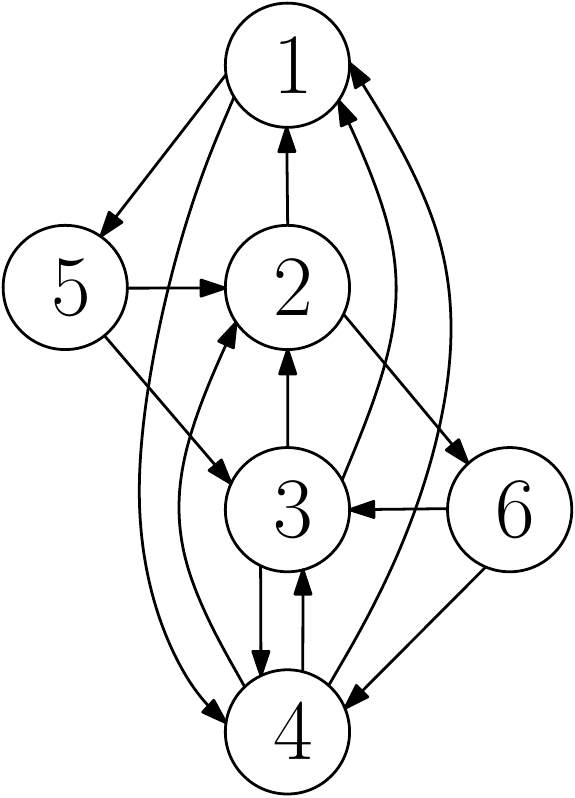}
		\label{4a}}
	\hspace{1cm}		
	\subfigure []{
		\includegraphics[height=3.2cm,keepaspectratio]{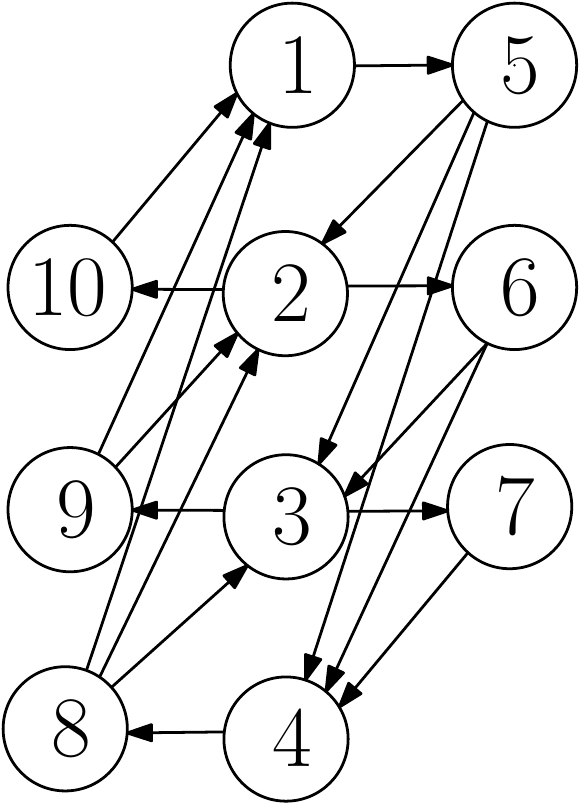}
		\label{4b}}
	\vskip-2pt
	\caption{(a) A $ 4 $-$\mathsf{GIC} $ digraph with six vertices, and (b) a $ 4 $-$\mathsf{GIC} $ digraph with ten vertices.}
	\label{optimalgraph}
	\vskip-10pt
\end{figure}

For $ K>3 $, a digraph $ D_K $ of this class is not an $  \mathsf{ICC} $ digraph, and its complement digraph (whose underlying graph is a complete graph) has chromatic number $ \goodchi (\bar{D}_K)=N=3K-2=\ell_{\mathsf{CL}} (D_K)$, and local chromatic number $ \goodchi_{\ell}(\bar{D}_K)=N-1=3K-3=\ell_{\mathsf{LC}}(D_K) $. Also, the digraph $ D_K$ has $\ell_{\mathsf{GICC}}(D_K)=\frac{2N+1}{3}=2K-1 $, and $\ell_{\mathsf{ICC}}(D_K) 
=\ell_{\mathsf{GICC}}(D_K)+(\lceil \frac{K}{3}\rceil -1)=2K+\lceil \frac{K}{3}\rceil -2 $. The gap, $ \ell_{\mathsf{LC}}(D_K)-\ell_{\mathsf{GICC}}(D_K)=K-2$, so for $ K>2 $, the gap grows linearly with $ K $. Similarly, $ \ell_{\mathsf{CL}}(D_K)- \ell_{\mathsf{GICC}}(D_K)=K-1$. Fig.~\ref{4b} depicts an example of a digraph in  this class with $ K=4 $.

\section{Conclusion}
In this work, we first defined a structure, called generalized interlinked cycle ($\mathsf{GIC} $), in directed graphs. A $\mathsf{GIC} $ consists of cycles that are interlinked in some manner (i.e., not disjoint). We then proposed a simple encoding scheme called the generalized interlinked cycle cover ($\mathsf{GICC} $). This scheme exploits $\mathsf{GICs} $ in the digraph, and generalizes the interlinked cycle cover ($ \mathsf{ICC} $) scheme, the clique cover scheme, and the cycle cover scheme. Furthermore, we characterized a class of digraphs where the $ \mathsf{GICC} $ is optimal (over all codes, including non-linear index codes), and showed that the $ \mathsf{GICC} $ scheme can outperform existing techniques (including partial clique cover, local chromatic number, composite-coding, and interlinked cycle cover).     

	\section{Acknowledgment}
	The authors would like to acknowledge Fatemeh Arbabjolfaei from the University of California, San Diego for providing the composite-coding upper bound for the digraph in Fig.~\ref{4a}.
	
	\bibliographystyle{IEEEtran}

\begin{thebibliography}{}
\providecommand{\url}[1]{#1}
\csname url@samestyle\endcsname
\providecommand{\newblock}{\relax}
\providecommand{\bibinfo}[2]{#2}
\providecommand{\BIBentrySTDinterwordspacing}{\spaceskip=0pt\relax}
\providecommand{\BIBentryALTinterwordstretchfactor}{4}
\providecommand{\BIBentryALTinterwordspacing}{\spaceskip=\fontdimen2\font plus
\BIBentryALTinterwordstretchfactor\fontdimen3\font minus
  \fontdimen4\font\relax}
\providecommand{\BIBforeignlanguage}[2]{{%
\expandafter\ifx\csname l@#1\endcsname\relax
\typeout{** WARNING: IEEEtran.bst: No hyphenation pattern has been}%
\typeout{** loaded for the language `#1'. Using the pattern for}%
\typeout{** the default language instead.}%
\else
\language=\csname l@#1\endcsname
\fi
#2}}
\providecommand{\BIBdecl}{\relax}
\BIBdecl

\end{thebibliography}


\begin{thebibliography}{10}
		\providecommand{\url}[1]{#1}
		\csname url@samestyle\endcsname
		\providecommand{\newblock}{\relax}
		\providecommand{\bibinfo}[2]{#2}
		\providecommand{\BIBentrySTDinterwordspacing}{\spaceskip=0pt\relax}
		\providecommand{\BIBentryALTinterwordstretchfactor}{4}
		\providecommand{\BIBentryALTinterwordspacing}{\spaceskip=\fontdimen2\font plus
			\BIBentryALTinterwordstretchfactor\fontdimen3\font minus
			\fontdimen4\font\relax}
		\providecommand{\BIBforeignlanguage}[2]{{%
				\expandafter\ifx\csname l@#1\endcsname\relax
				\typeout{** WARNING: IEEEtran.bst: No hyphenation pattern has been}%
				\typeout{** loaded for the language `#1'. Using the pattern for}%
				\typeout{** the default language instead.}%
				\else
				\language=\csname l@#1\endcsname
				\fi
				#2}}
		\providecommand{\BIBdecl}{\relax}
		\BIBdecl
		
		\bibitem{ISCOD}
		Y.~Birk and T.~Kol, ``Informed-source coding-on-demand {(ISCOD)} over broadcast
		channels,'' in \emph{Proc. {IEEE} {INFOCOM}}, vol.~3, San Francisco, CA, Mar.
		1998, pp. 1257--1264.
		
		\bibitem{maisbound}
		Z.~Bar-Yossef, Y.~Birk, T.~S. Jayram, and T.~Kol, ``Index coding with side
		information,'' \emph{{IEEE} Transactions on Information Theory}, vol.~57,
		no.~3, pp. 1479--1494, Mar 2011.
		
		\bibitem{linearcodeoptimal}
		L.~Ong, ``Linear codes are optimal for index-coding instances with five or
		fewer receivers,'' in \emph{Proc. {IEEE} International Symposium on
			Information Theory (ISIT)}, June 2014, pp. 491--495.
		
		\bibitem{chaudhary}
		M.~A.~R. Chaudhry, Z.~Asad, A.~Sprintson, and M.~Langberg, ``On the
		complementary index coding problem,'' in \emph{Proc. {IEEE} International
			Symposium on Information Theory (ISIT)}, July 2011, pp. 224--248.
		
		\bibitem{eden}
		E.~Chlamtac and I.~Haviv, ``Linear index coding via semidefinite programming,''
		in \emph{Proc. 23rd Annu. ACM-SIAM Symp. on Discrete Algorithms (SODA)},
		2012, pp. 406--419.
		
		\bibitem{NPhard1}
		R.~Peeters, ``Orthogonal representaions over finite fields and the chromatic
		number of graphs,'' \emph{Combinatorica}, vol.~16, no.~3, pp. 417--431, Sept
		1996.
		
		\bibitem{neely}
		\BIBentryALTinterwordspacing
		M.~J. Neely, A.~S. Tehrani, and Z.~Zhang, ``Dynamic index coding for wireless
		broadcast networks,'' \emph{{IEEE} Transactions on Information Theory}, vol.~59,
		no.~11, pp. 7525--7540, Nov 2013.
		\BIBentrySTDinterwordspacing
		
		\bibitem{linearprogramming}
		\BIBentryALTinterwordspacing
		A.~Blasiak, R.~D. Kleinberg, and E.~Lubetzky, ``Broadcasting with side information: Bounding and approximating the broadcast rate," \emph{{IEEE} Transactions on Information Theory}, vol.~59,
		no.~9, pp. 5811--5823, Sept 2013.
		\BIBentrySTDinterwordspacing
		
		\bibitem{localgarphcoloring}
		K.~Shanmugam, A.~G. Dimakis, and M.~Langberg, ``Local graph coloring and index
		coding,'' in \emph{Proc. {IEEE} International Symposium on Information Theory
			(ISIT)}, July 2013, pp. 1152--1156.
		
		\bibitem{ourpaper}
		\BIBentryALTinterwordspacing
		C.~Thapa, L.~Ong, and S.~J. Johnson, ``A new index coding scheme exploiting
		interlinked cycles,'' in \emph{Proc. {IEEE} International Symposium on Information Theory (ISIT)},
		June 2015, pp. 1024--1028. 
		\BIBentrySTDinterwordspacing
		
		\bibitem{composite}
		F.~Arbabjolfaei, B.~Bandemer, Y.-H. Kim, E.~Sasoglu, and L.~Wang, ``On the
		capacity region for index coding,'' in \emph{Proc. {IEEE} International
			Symposium on Information Theory (ISIT)}, July 2013, pp. 962--966.
		
		\bibitem{localtimesharing}
		F.~Arbabjolfaei and Y.-H. Kim, ``Local time sharing for index coding,'' in
		\emph{Proc. {IEEE} International Symposium on Information Theory (ISIT)},
		June 2014, pp. 286--290.
		
	\end{thebibliography}
	
	\appendices
	\section{Proof of Proposition \ref{lemma3}} \label{append1}	
		A $ K $-$ \mathsf{GIC} $ sub-digraph $ D_K $ has some properties captured in the following lemmas, which we will use to prove $N^+_{T_i}(v)=N^+_{D_K}(v) $. Here we consider $ T_i $ and $ T_j $ as any two distinct trees present in $ D_K $.  
		\begin{Lemma} \label{lemma1}
			If a vertex $ v\in V(T_i)$, $ v\in V(T_j) $, and $ v\notin V_{\mathrm{I}} $, then the set of leaf vertices that fan out from the common vertex $ v $ in each tree is a subset of $ V_{\mathrm{I}}\setminus \{i,j\} $. 
		\end{Lemma}
		\begin{IEEEproof}
			In a tree $ T_i $ (see Fig. \ref{fig3}), for any vertex $ v\in V (T_i) $ and $ v\notin V_{\mathrm{I}} $, let $ L_{T_i} (v)$ be a set of leaf vertices that fan out from the vertex $ v $. If the vertex $ j\in L_{T_i} (v) $, then there exists a path from vertex $ v $ to $ j $ in the tree $ T_i $. However, in the tree $ T_j $, there is a path from vertex $ j $ to $ v $. Thus in the sub-digraph\footnote{As $ D_K=\bigcup_{\forall i \in V_{\mathrm{I}}} T_i $, a path present in any $ T_i $ also present in $ D_K $.} $ D_K $, we obtain a path from vertex $ v $ to $ j $ (via $ T_i $) and vice versa (via $ T_j $). As a result, an $ I $-cycle containing $ j $ is present. This contradicts the condition 1 (i.e., no $ I $-cycle) for a $ D_K $. Hence, $ j\notin L_{T_i} (v) $. In other words, $ L_{T_i} (v) \subseteq V_{\mathrm{I}} \setminus \{i,j\} $. Similarly, $ L_{T_j} (v) \subseteq V_{\mathrm{I}} \setminus \{i,j\} $. 
		\end{IEEEproof}
		
		\begin{Lemma} \label{lemma2}
			If a vertex $ v\in V(T_i)$, $ v\in V(T_j) $, and $ v\notin V_{\mathrm{I}} $, then the out-neighborhood of the vertex $ v $ is same in both of the trees, i.e., $ N^+_{T_i}(v)=N^+_{T_j}(v) $. 
		\end{Lemma}
		\begin{IEEEproof}            
			Here the proof is done by contradiction. Let us suppose that $ N^+_{T_i}(v)\neq N^+_{T_{j}}(v) $. 
			
			For this proof we refer to Fig. \ref{fig3}. This proof has two parts. In the first part, we prove $ L_{T_i}(v)=L_{T_j}(v)$, and then prove $N^+_{T_i}(v)=N^+_{T_j}(v) $ in the second part.
			
			From Lemma \ref{lemma1}, $ L_{T_i} (v) $ is a subset of $ V_{\mathrm{I}} \setminus \{i,j\} $. Now pick a vertex $ c $ belongs to $ V_{\mathrm{I}} \setminus \{i,j\} $ such that $ c \in L_{T_i} (v)$ but $ c\notin L_{T_j}(v) $ (such $ c $ exists since we suppose that $ L_{T_i}(v)\neq L_{T_{j}}(v) $). In tree $ T_i $, there exists a directed path from the vertex $ i $ which includes the vertex $ v $, and ends at the leaf vertex $ c $. Let this path be $ P_{i\rightarrow c}(T_i)$. Similarly, in tree $ T_j $, there exists a directed path from the vertex $ j $, which doesn't include the vertex $ v $ (since $ c\notin L_{T_j}(v) $), and ends at the leaf vertex $ c $. Let this path be $ P_{j\rightarrow c}(T_j) $. However, in the digraph $ D_K $, we can also obtain a directed path from the vertex $ j $ which passes through the vertex $ v $ (via $ T_j$), and ends at the leaf vertex $ c $ (via $ T_i $). Let this path be $ P_{j\rightarrow c}(D_K) $. The paths $ P_{j\rightarrow c}(T_j) $ and $ P_{j\rightarrow c}(D_K)$ are different which indicates the existence of multiple $ P $-paths from the vertex $ j $ to $ c $ in $ D_K $, this contradict the condition 2 for a $ D_K $. Consequently, $ L_{T_i}(v)=L_{T_j}(v)$.
			\begin{figure}[t] 
				\centering
				\includegraphics[height=4.4cm,keepaspectratio]{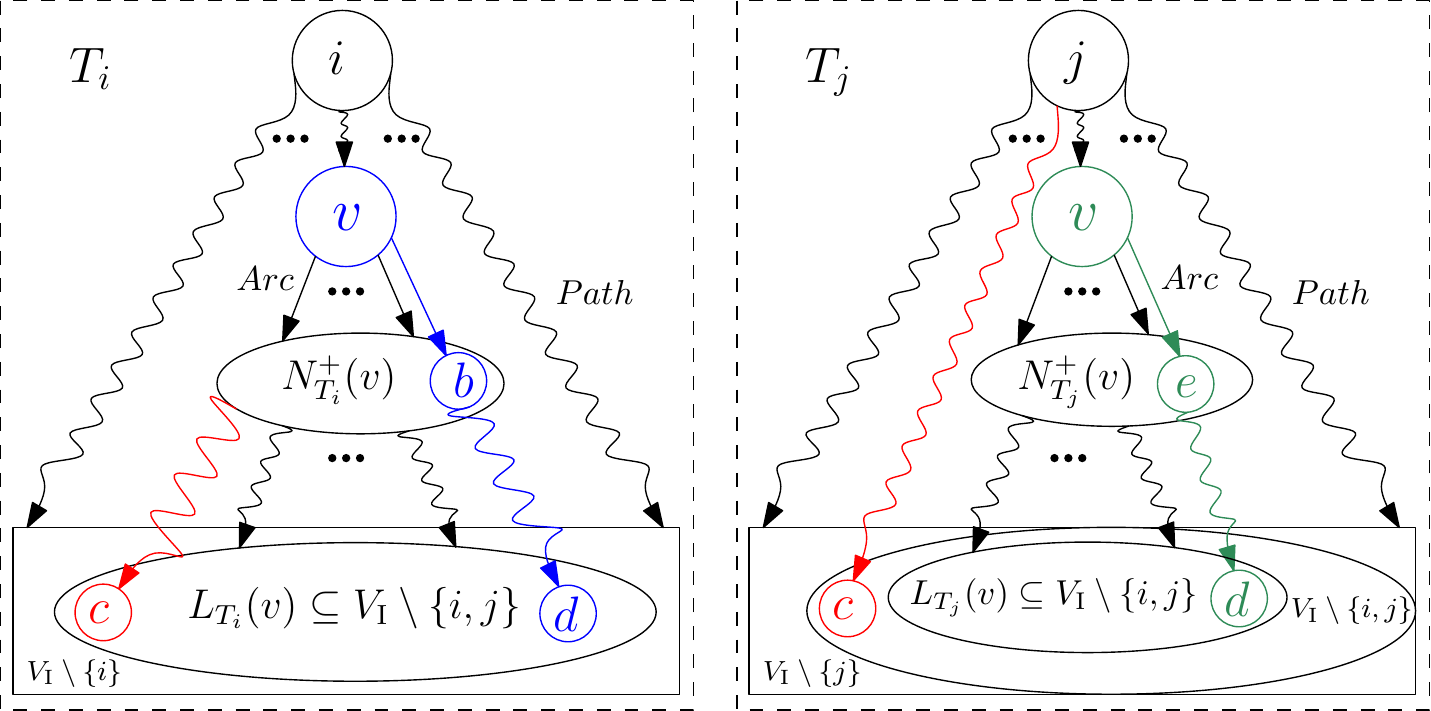}        
				\caption{Directed rooted trees $ T_i $ and $ T_j $ with roots $ i $ and $ j $ respectively, and a non inner vertex $ v $ in common. Here we have used straight arrow to indicate an arc, and curly arrow to indicate a path.}  
				\label{fig3}
			\end{figure}
						
			Now we pick a vertex $ b $ such that, without loss of generality, $b\in N^+_{T_i} (v) $ but $b\notin N^+_{T_j} (v) $ (such $ b $ exists since we assumed that $ N^+_{T_i}(v)\neq N^+_{T_{j}}(v) $). Furthermore, we have two cases for $ b $, which are (case 1) $ b\in L_{T_i}(v)$, and (case 2) $ b\notin L_{T_i}(v)$. Case 1 is addressed in the first part of this proof. On the other hand, for case 2, we pick a leaf vertex $ d \in L_{T_i}(b)$ such that there exists a path (see Fig. \ref{fig3}) that starts from $ v $ followed by $ b $, and ends at $ d $, i.e., $ \langle v,b,\dotsc,d \rangle $ exists in $ T_i $. A path $ \langle j,\dotsc,v \rangle $ exists in $ T_j $. Thus a path $ \langle j,\dotsc,v,b,\dotsc,d \rangle $ exists in $ D_K $. From the first part of the proof, we have $ L_{T_i}(v)=L_{T_j}(v)$, so $ d\in L_{T_j}(v) $. Now in $ T_j $, there exists a path from $ j $ to $ d $, which includes vertex $ v $ followed by a vertex $ e$ such that $ e\in N^+_{T_j} (v)$ and $e\neq b$ (as $ b\notin N^+_{T_j} (v) $), and the path ends at $ d $, i.e., $ \langle j,\dotsc,v,e,\dotsc,d \rangle $ which is different from $ \langle j,\dotsc,v,b,\dotsc,d \rangle $. So multiple $ P $-paths are observed at $ d $ from $ j $. This contradicts condition 2 for a $ D_K $. Consequently, $ N^+_{T_i}(v)= N^+_{T_j}(v) $.  
		\end{IEEEproof} 
	
		\begin{IEEEproof} [Proof of Proposition \ref{lemma3}]
			For any $ v \in V(T_i) $ from Lemma~\ref{lemma2}, $N^+_{T_i}(v)=N^+_{T_j}(v) $ for all $ \{j: v \in T_j\} $. Since $ D_K=\bigcup_{\forall i \in V_{\mathrm{I}}} T_i $, vertex $ v $ must have the same out-neighborhood in $ D_K $ as well. 
		\end{IEEEproof}
		
	\section{} \label{append2}	
	In this section, for the tree $ T_i $, we mathematically compute $ Z_i $ and $ w_{\mathrm{I}}\oplus Z_i $. These results are referred in Section \ref{secb}.
	
	In the tree $ T_i $, 
	\begin{align} \label{eqC}
	Z_i&=\bigoplus \limits_{j\in V(T_i)\setminus V_{\mathrm{I}} } w_j  \nonumber \\
	&=\bigoplus \limits_{j\in V(T_i)\setminus V_{\mathrm{I}}} \left( x_j \oplus \bigoplus \limits_{q\in N^+_{D_K}(j)}x_q \right) \nonumber \\
	&= \bigoplus \limits_{j\in V(T_i)\setminus V_{\mathrm{I}}}\left( x_j \oplus \bigoplus \limits_{q\in N^+_{D_K}(j)\setminus V_{\mathrm{I}}} x_q  \oplus \bigoplus \limits_{ q\in N^+_{D_K}(j)\cap V_{\mathrm{I}}} x_q   \right)\nonumber \\
	&= X_{V(T_i)} \oplus X'_{V(T_i)}.  
	\end{align}
	
	\begin{figure*} [!t]
		\begin{multline*}  \label{eqA}
		X_{V(T_i)}= 
		\bigoplus \limits_{j_1\in N^+_{D_K}(i)\setminus V_{\mathrm{I}}}
		\left[ 
		\left(
		x_{j_1} 
		\oplus 
		\textcolor{blue}{
			\bigoplus \limits_{q\in N^+_{D_K}(j_1)\setminus V_{\mathrm{I}}} 
			x_{q_1} } 
		\right) 
		\oplus 
		\textcolor{blue}{
			\bigoplus \limits_{j_2\in N^+_{D_K}(j_1)\setminus V_{\mathrm{I}}} }
		\left[ 
		\left(
		\textcolor{blue}{
			x_{j_2} }
		\oplus 
		\textcolor{ao}{
			\bigoplus \limits_{q\in N^+_{D_K}(j_2)\setminus V_{\mathrm{I}}}
			x_{q_2} } 
		\right) 
		\textcolor{ao}{\oplus} \dotsc \textcolor{copper}{\oplus }  \right. \right. \\	
		\textcolor{copper}{
			\bigoplus \limits_{j_{H-2}\in N^+_{D_K}(j_{H-3})\setminus V_{\mathrm{I}}} }
		\left[ \hskip-2pt
		\left(
		\textcolor{copper}{
			x_{j_{H-2}} }
		\oplus \hskip-2pt
		\textcolor{red}{
			\bigoplus \limits_{q\in N^+_{D_K}(j_{H-2})\setminus V_{\mathrm{I}}} 
		\hskip-2pt	x_{q_{H-2}}  }
		\right) 
		\oplus  
		\left. \left. \left. 
		\textcolor{red}{	 
			\bigoplus \limits_{\substack{j_{H-1}\in \\ N^+_{D_K}(j_{H-2})\setminus V_{\mathrm{I}}}} } \hskip-3pt
		\left[ 
		\left(
		\textcolor{red} {
			x_{j_{H-1}} }
		\oplus 
		\bigoplus \limits_{\tunderbrace{q\in N^+_{D_K}(j_{H-1})\setminus V_{\mathrm{I}}}_{=0}} \hskip-5pt
		x_{q_{H-1}}
		\right.  
		\right) 
		\right]
		\right] \dotsc 
		\right] 
		\right] 
	\end{multline*}
	\begin{equation} \label{eqB}
	=\bigoplus \limits_{j\in N^+_{D_K}(i)\setminus V_{\mathrm{I}}} x_{j}.  \hspace{12.35cm}
	\end{equation}
	\begin{align} \label{xvt2}
	& w_{\mathrm{I}} \oplus Z_i \nonumber \\
	&=x_i\oplus  \bigoplus \limits_{j\in V_{\mathrm{I}}\setminus \{i\}}   x_j 
	\oplus 
	\bigoplus \limits_{k\in N^+_{D_K}(i)\setminus V_{\mathrm{I}}} x_{k} 
	\oplus 
	\bigoplus \limits_{\substack{q:q\in V_{\mathrm{I}}\setminus \{i\}\\ \text{\&}\ q \notin N^+_{D_K}(i)}}  x_q  \nonumber \hspace{5cm}\\
	&=x_i
	\oplus
	\left( 
	\bigoplus \limits_{\substack{j:j\in V_{\mathrm{I}}\setminus \{i\}\\ \text{\&}\ j \in N^+_{D_K}(i)}} \hskip-2pt  x_{j}
	\oplus
	\textcolor{red}{
	\bigoplus \limits_{\substack{j:j\in V_{\mathrm{I}}\setminus \{i\}\\ \text{\&}\ j \notin N^+_{D_K}(i)}} x_{j} }	
	\right) 
	\oplus \hskip-2pt
	\bigoplus \limits_{k\in N^+_{D_K}(i)\setminus V_{\mathrm{I}}}\hskip-2pt x_{k}	
	\oplus
	\left( \textcolor {red}{
	\bigoplus \limits_{\substack{q:q\in V_{\mathrm{I}}\setminus \{i\}\\ \text{\&}\ q \notin N^+_{D_K}(i)}}\hskip-2pt x_{q} }
	\right) 
	=x_i
	\oplus
	\left( \hskip-2pt
	\bigoplus \limits_{\substack{j:j\in V_{\mathrm{I}}\setminus \{i\}\\ \text{\&}\ j \in N^+_{D_K}(i)}} \hskip-2pt x_{j}
	\oplus \hskip-2pt
	\bigoplus \limits_{k\in N^+_{D_K}(i)\setminus V_{\mathrm{I}}}\hskip-2pt x_{k} 
	\right). 
	\end{align}
		\hrulefill
	\end{figure*}
		
	Where,
	$ X_{V(T_i)}\triangleq \bigoplus \limits_{j\in V(T_i)\setminus V_{\mathrm{I}}}\left(x_j \oplus \bigoplus \limits_{q\in N^+_{D_K}(j)\setminus V_{\mathrm{I}}}x_q  \right) $, and $ X'_{V(T_i)}\triangleq \bigoplus \limits_{j\in V(T_i)\setminus V_{\mathrm{I}}}\left( \bigoplus \limits_{q\in N^+_{D_K}(j)\cap V_{\mathrm{I}}}x_q \right)=\bigoplus \limits_{\substack{q:q\in V_{\mathrm{I}}\setminus \{i\}\\ \text{\&}\ q \notin N^+_{D_K}(i)}} x_{q}$.

	Here $ X'_{V(T_i)}$ is bitwise XOR of messages requested by all of the leaf vertices which branch from all of the non leaf and the non root vertices in $ T_i $. If we expand $ X_{V(T_i)} $ as per the group of vertices according to their depth, then the intermediate terms cancel out, and we get \eqref{eqB} (we have use the same color to indicate the terms that cancel out each other).
	
	Now substituting $ X_{V(T_i)} $ of \eqref{eqB} and $ X'_{V(T_i)} $ in \eqref{eqC}, we get
	\begin{align} \label{eqD}
	Z_i =\bigoplus \limits_{k\in N^+_{D_K}(i)\setminus V_{\mathrm{I}}} x_{k}	
	\oplus
	\left( 
	\bigoplus \limits_{\substack{q:q\in V_{\mathrm{I}}\setminus \{i\}\\ \text{\&}\ q \notin N^+_{D_K}(i)}} x_{q}
	\right).
	\end{align} 
	
	Now in the tree $ T_i $, $ w_{\mathrm{I}} \oplus Z_i $ yields \eqref{xvt2}, which is XOR of messages requested by the inner vertex $ i $, and its out-neighbourhood vertices.

\end{document}